\documentclass[aps,prl,twocolumn,showpacs,superscriptaddress,groupedaddress]{revtex4-1}  
\usepackage{graphicx}  
\usepackage{dcolumn}  
\usepackage{bm}        
\usepackage{amssymb}   
\usepackage{natbib}   
\usepackage{amsmath}
\usepackage{mathtools}
\usepackage{xfrac}
\usepackage{caption}
\usepackage{color}
\usepackage{soul}
\usepackage{hyperref}
\usepackage{float}
\usepackage{lipsum}
\usepackage{enumerate}
\begin{document}
\title{Short- and Long- Time Transport Structures in a Three Dimensional Time Dependent Flow}
\author{R. Chabreyrie}
\address{Department of Mechanical and Aerospace Engineering, Jacobs School of Engineering, UCSD, 9500 Gilman Drive, La Jolla CA 92093-0411, USA}
\author{S. G. Llewellyn Smith}
\address{Department of Mechanical and Aerospace Engineering, Jacobs School of Engineering, UCSD, 9500 Gilman Drive, La Jolla CA 92093-0411, USA}   
\begin{abstract}
Lagrangian transport structures for three-dimensional and time-dependent fluid flows are of great interest in numerous applications, particularly for geophysical or oceanic flows. In such flows, chaotic transport and mixing can play important environmental and ecological roles, for examples in pollution spills or plankton migration.
In such flows, where  simulations or observations are typically available only over a short time, understanding the difference between short-time and long-time transport structures is critical.
In  this paper,  we use a set of classical (i.e. Poincar\'e section, Lyapunov exponent) and alternative (i.e. finite time Lyapunov exponent, Lagrangian coherent structures) tools from dynamical systems theory that analyze chaotic transport both qualitatively  and quantitatively.
With this set of tools we are able to reveal, identify  and highlight differences between short- and long-time transport structures
inside a flow composed of a primary horizontal contra-rotating vortex chain,  small lateral oscillations and a weak Ekman pumping.
The difference is mainly the existence of regular or extremely slowly
developing chaotic regions that are only present at short time.
\end{abstract}

\maketitle
\section{Introduction}
Even if turbulence is not present in laminar flows, scalar transport in such systems can be very rich and complex.
In particular, it is well known that in such systems, transport through successive mechanisms of stretching and folding of material lines in multiple directions may occur. For the past three decades, the kinematics viewpoint of fluid transport, i.e. the behavior of advected/passive particle trajectories, has received much attention\cite{Aref:2002}.  The generation of  Lagrangian chaotic transport in a small Reynolds number flow is generally achieved by adding at least one degree of freedom to a two dimensional incompressible base flow. Such  degrees of freedom 
often take the form of a weak time dependence \cite{Solomon:1988a, Solomon:1988b, Paoletti:2006, Mancho:2006} and/or weak dependence on the third dimension \cite{Baj, KroujilineandStone:1999}.

The most general approach along these lines is to consider an incompressible three dimensional flow, the superposition of a two dimensional integrable flow ${\bm V}\left(x,y\right)$ and a small three dimensional unsteady perturbation $\epsilon_0{\bm V}\left(x,y,t\right)+\epsilon_1{\bm V}\left(x,y,z\right)$, $0\leq\epsilon_{0,1}\ll1$.
If  the perturbed flow is  unsteady and two-dimensional ($\epsilon_0\neq0, \epsilon_1=0$), or steady and three-dimensional ($\epsilon_0=0, \epsilon_1\neq0$),
the kinematic behavior and Lagrangian transport structure are well known and established, with notably the existence of two-dimensional Kolmogorov Arnold Moser (KAM) tori acting as barrier to transport.
   
If the perturbed flow is composed of  both an unsteady two dimensional and a steady three-dimensional perturbation ($\epsilon_0\neq0, \epsilon_1\neq0$), the situation is more complex with no theory still completely established yet. 
Depending on the number of fast (i.e.~action) and slow (i.e.~angle) variables that are necessary to described the system, the kinematic behavior, and consequently the Lagrangian transport structure, can be very different.

On the one hand, in flows described by one slow and two fast variables, action-angle-angle flows, the kinematic and Lagrangian space structure is similar to the simpler case (i.e.~unsteady two dimensional or steady three-dimensional) with the presence of KAM-like regular tori acting as an impermeable barrier to transport.
On the other hand, in action-action-angle flows, Arnold-like diffusion may appear, enabling essentially complete mixing via resonance phenomena.

These action-action-angle flows have an important place in the studies of Lagrangian transport. Indeed, such flows arise frequently in small (e.g.~microfluidic devices) and large (e.g.~geophysical or oceanic flows) scales  where geometric symmetries severely constrain the flow structure, leading to the apparition of multiple flow actions.

While many previous works have focus on describing, both qualitatively \cite{CambridgeJournals:379074, springerlink:10.1007/BF01026490, PhysRevLett611799, PhysRevLett753669} and quantitatively \cite{VNM:2006, VWG:2007, Neishtadt:05, PRE780263}, the long-time transport structure of these perturbed action-action-angle flows, we turn our attention to the short-time behavior. One motivation is that studying the long-time behavior of the system requires  knowing or modeling the system over long times,  which is not always possible or physically realistic. For example in oceanic flows simulations or observations are typically available only during a short time. Specifically, we seek to characterize and identify the transport structures present at short time with the help of alternative tools.
 
This paper is organized as follows. The physical model, as well as its assumptions and corresponding dynamical system, are first described.
Then  the different transport characteristics possible in this system are  summarized.
The Lagrangian structures at long time via Poincar\'e sections and infinite time Lyapunov exponent maps are shown. Finally, we present both qualitatively and quantitatively,  the evidence of  hidden  transport structures present only at short time. 
Then these short time transport structures are revealed and clearly identified as Lagrangian coherent structures.
\section{\label{Sec:Toy_Model}Toy model}
\subsection{Apparatus}
\begin{figure*}[ht]
\begin{center}
\includegraphics[scale=.70]{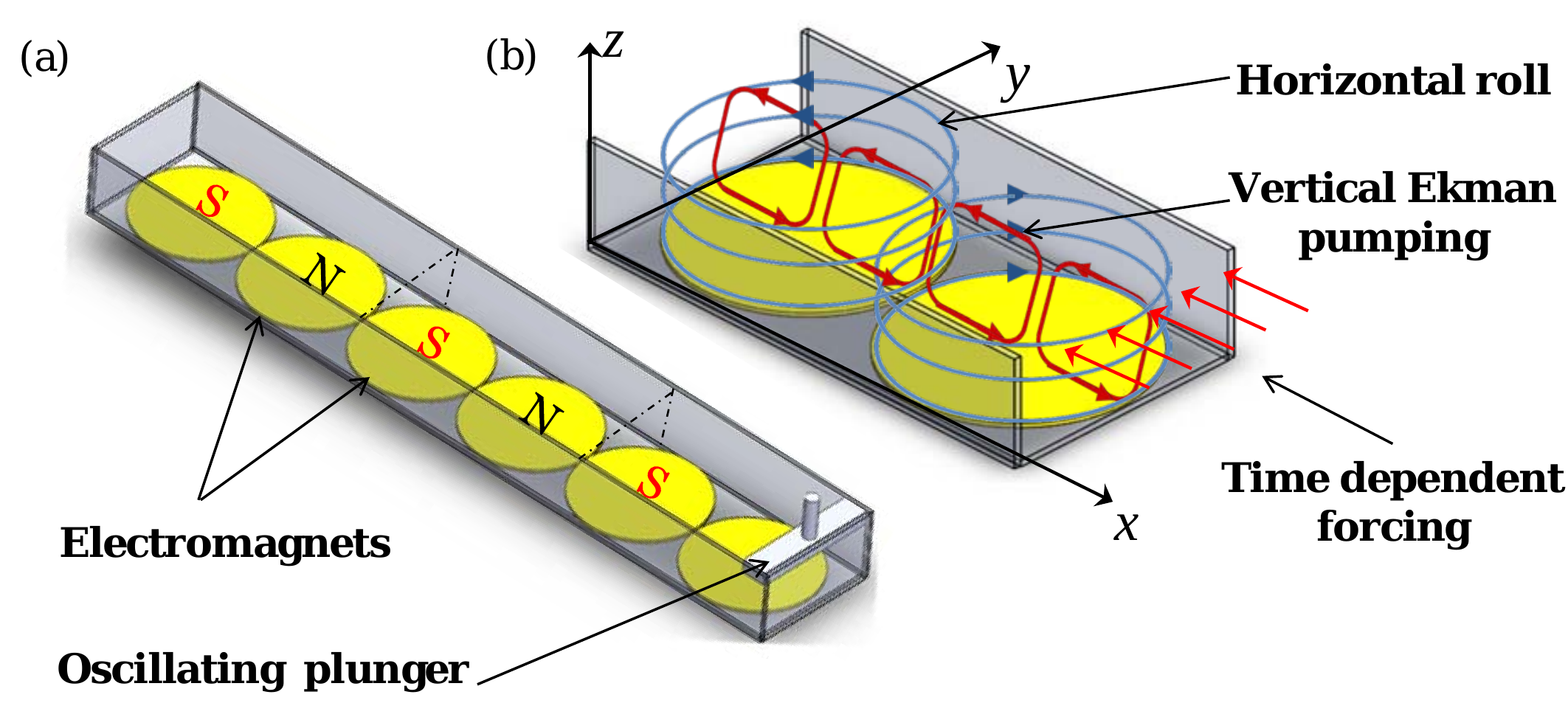}
\end{center}
\caption{a) Sketch of the experimental device. A long channel characterized by a chain of electromagnets embedded into the floor with an oscillating plunger attached to one of the sides. b) Fluid flow: the base flow (blue) is a chain of alternating horizontal vortices; a weak perturbative flow (red) is generated naturally by Ekman pumping, which draws fluid inward along the bottom of the vortices and up the vortex centers.}
\label{fig1}
\end{figure*}

We consider the same magneto-hydrodynamic (MHD) flow system as \citeauthor{Solomon:1988a} (see Fig.~\ref{fig1}). An electric current
passes through a thin layer of an electrolytic solution (salt water or dilute $H_2S0_4$) contained
in a rectangular channel. The current interacts with the magnetic field produced by a series of
magnets of alternating polarity embedded into the bottom wall, resulting in a periodic array of
vortices separated by separatrices directly above the centers of the magnets. Time-dependent
forcing is introduced in the system by displacing the fluid slowly back and forth across the
magnets with the use of a plunger. Typical frequencies of the forcing are around 0.050\ensuremath{{\rm~Hz}},
corresponding to a period significantly longer than the viscous time scale (of the order of 4\ensuremath{{\rm~Hz}}).


\subsection{Flow}
 Such an apparatus produces a flow composed of a primary horizontal alternating vortex chain (produced by the electromagnets), small lateral oscillations (created by the plunger) and a  weak vertical secondary flow (generated by Ekman pumping). 
The features of the flow in Fig.~\ref{fig1}, i.e.~the crisscrossing of swirling rolls and time-dependent oscillations, are generic features in chaotic mixing problems and are present in many flows, notably in the ocean. 
A two-dimensional pair of contra-rotating swirling rolls  is often used   to model oceanic double gyres produced by the wind over the ocean's surface \cite{aharon:056603, WigginsLTGJW, Samelson, Zhengyu, Lekien:2007, Yang:1994}, while the addition of a time dependent forcing and a vertical velocity can be justified to simulate  wind variations or tidal effects and Ekman pumping present near fronts or induced by night convection \cite{Mahadevan2006241}.
Critically, a simple analytical expression for the  velocity field  has been experimentally verified using a relatively easy to build small-scale apparatus.
\subsection{Velocity field}
We start from the Eulerian velocity field ${\bm V}=(u,v,w)$ and consider the particle path equations
\begin{eqnarray}
\label{VF1}
u=&\frac{dx}{dt}&=-\cos(x+\epsilon_0\sin\omega t)\sin y + \epsilon_1\sin(2x+2\epsilon_0\sin\omega t)\sin z,\\
\label{VF2}
v=&\frac{dy}{dt}&= \sin(x+\epsilon_0\sin\omega t)\cos y + \epsilon_1\sin(2y)\sin z,\\
\label{VF3}
w=&\frac{dz}{dt}&=2\epsilon_1 \cos z [\cos(2x+2\epsilon_0\sin\omega t)+\cos(2y)],
\end{eqnarray}
where $\omega$,  $\epsilon_0$ and $\epsilon_1$ stand for the base oscillation's frequency, its amplitude and the Ekman pumping strength, respectively.

\begin{figure*}[ht]
\begin{center}
\includegraphics[scale=.8]{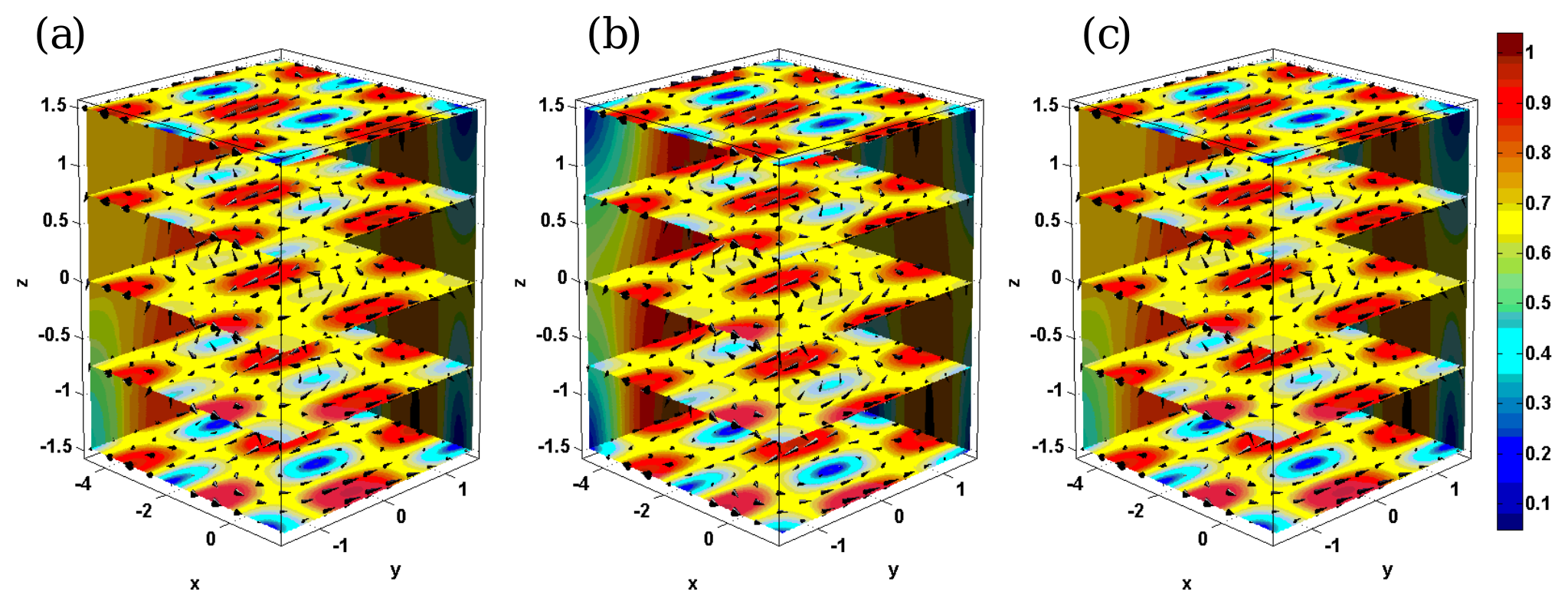}
\end{center}
\caption{ a--c) Velocity field and contour levels of the velocity intensity inside a generic cell for the three-dimensional time dependent case at time $t=\pi/2\omega,\pi\omega,3\pi/4\omega$.}
\label{fig2}
\end{figure*}

The resulting velocity field  and contour levels of the velocity magnitude in a generic cell for the three dimensional time-dependent case (i.e.~$\epsilon_0 \neq 0$ and $\epsilon_1 \neq 0$) are represented in Fig.~\ref{fig2}(a--c). Figures~\ref{fig2} illustrates how, inside a cell, the vortex pair is horizontally pushed forward and backward in time by the action of the plunger and how this two contra-rotating rolls structure is perturbed in third dimension by the vertical Ekman pumping.

\section{\label{Sec:Characteristics} Time scales  and transport characteristics as a function of parameters}
\subsection{Non-chaotic case}
\begin{figure}[ht]
\begin{center}
\includegraphics[scale=.70]{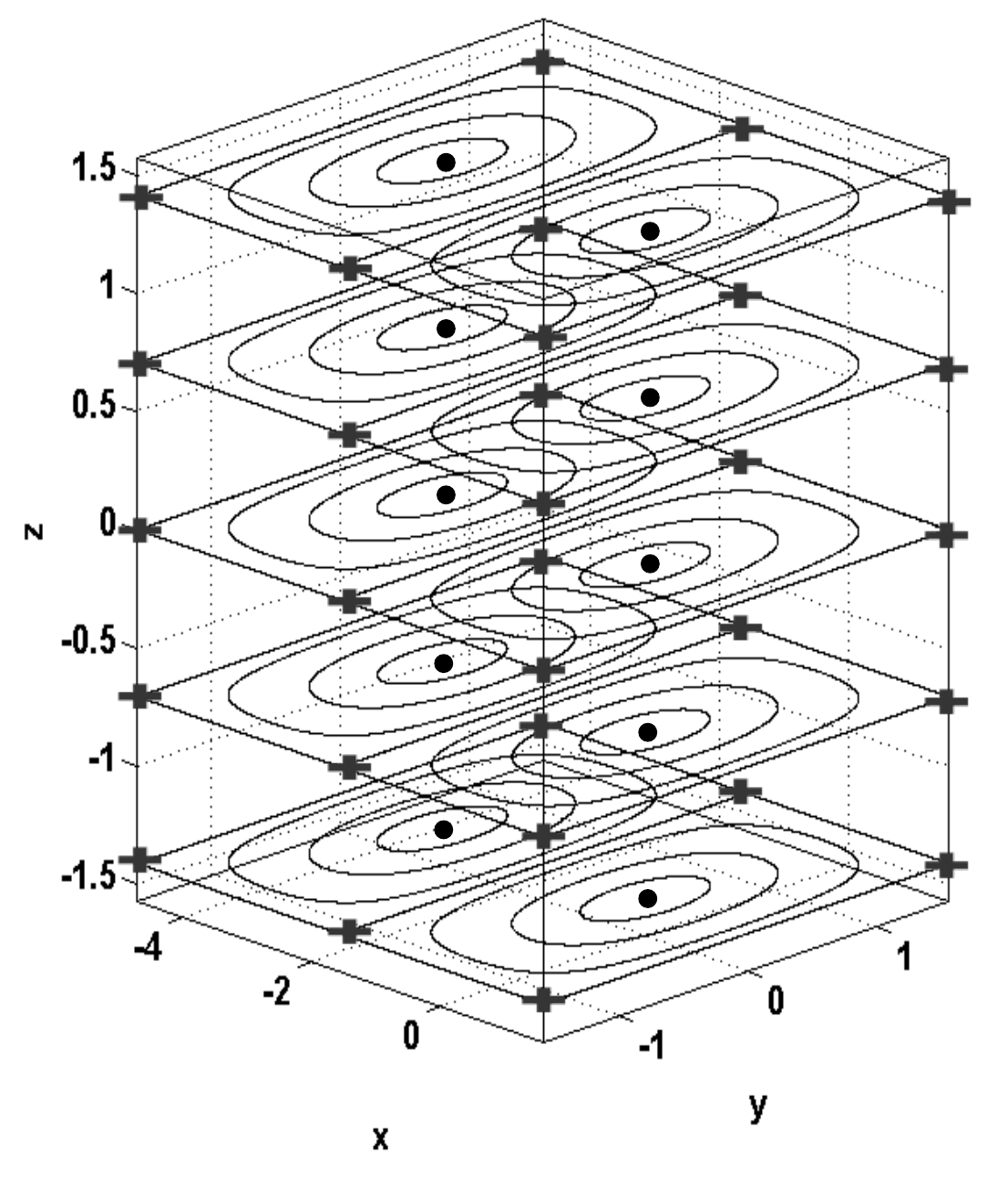}
\end{center}
\caption{Streamlines in the integrable case $\epsilon_{0,1}=0$, degenerate elliptic points (bold points) and hyperbolic points (bold crosses).}
\label{fig11}
\end{figure}

The steady two-dimensional case, corresponding to $\epsilon_{0,1}=0$, i.e.~without periodic forcing and Ekman pumping, is characterized by two invariants (time-independent quantities), the stream function $\psi_0=\sin x \sin y~\in\left[-1,1\right]$ and the vertical position $z \in \left[-\pi/2,\pi/2\right]$.
The streamlines of
the non-chaotic case are curves of constant $\psi$ and $z$. The flow consists of a series of periodic cells, each cell consisting of two side-by-side recirculating rolls rotating in opposite directions as represented in Fig.~\ref{fig11}. 
Within a generic cell, pathlines are organized into two sets of  closed lines around a vertical line of degenerated elliptic fixed points, located at $\left(x=-\pi,y=0\right)$ and $\left(x=0,y=0\right)$.  
In addition, heteroclinic pathlines, connect degenerated hyperbolic fixed points located vertically on the corners of a generic half-cell, i.e.~$\left(x=-3\pi/2,y=\pm \pi/2\right)$, $\left(x=-\pi,y=\pm \pi/2\right)$, $\left(x=\pi/2,y=\pm \pi/2\right)$ (see Fig.~\ref{fig11}).\\
The frequency of motion on a
streamline  is  given by
\begin{equation*}
\frac{\pi}{\Omega(\psi)} = \int^{x_{max}}_{-\pi/2}\frac{dx}{\sqrt{\cos^2x-\psi^2}},
\end{equation*}
with $x_{max}=1/\cos\psi$ (see Ref.~\citep{PRE780263} for more details).
On every streamline we introduce a uniform phase $\chi  \bmod 2\pi$
such that $\chi=0$ on the ${\bm e}_{x}-{\bm e}_{z}$ plane and
$\dot{\chi}=\Omega\left(\psi\right)$. With this uniform phase, the non-chaotic
flow can be described by using the variables $(\psi,z,\chi)$ instead of the
Cartesian coordinates $(x,y,z)$, with
\begin{eqnarray*}
\dot{\psi}&=&0,\\
\dot{z}&=&0, \\
\dot{\chi}&=&\Omega(\psi).
\end{eqnarray*}
Such a system is generally classified as action-action-angle, where the two invariants $\psi$ and $z$ are the two
actions (constant quantities) and $\chi$ is the angle (oscillating quantity).
\subsection{Two-dimensional chaotic case}
\begin{figure}[ht]
\begin{center}
\includegraphics[scale=.70]{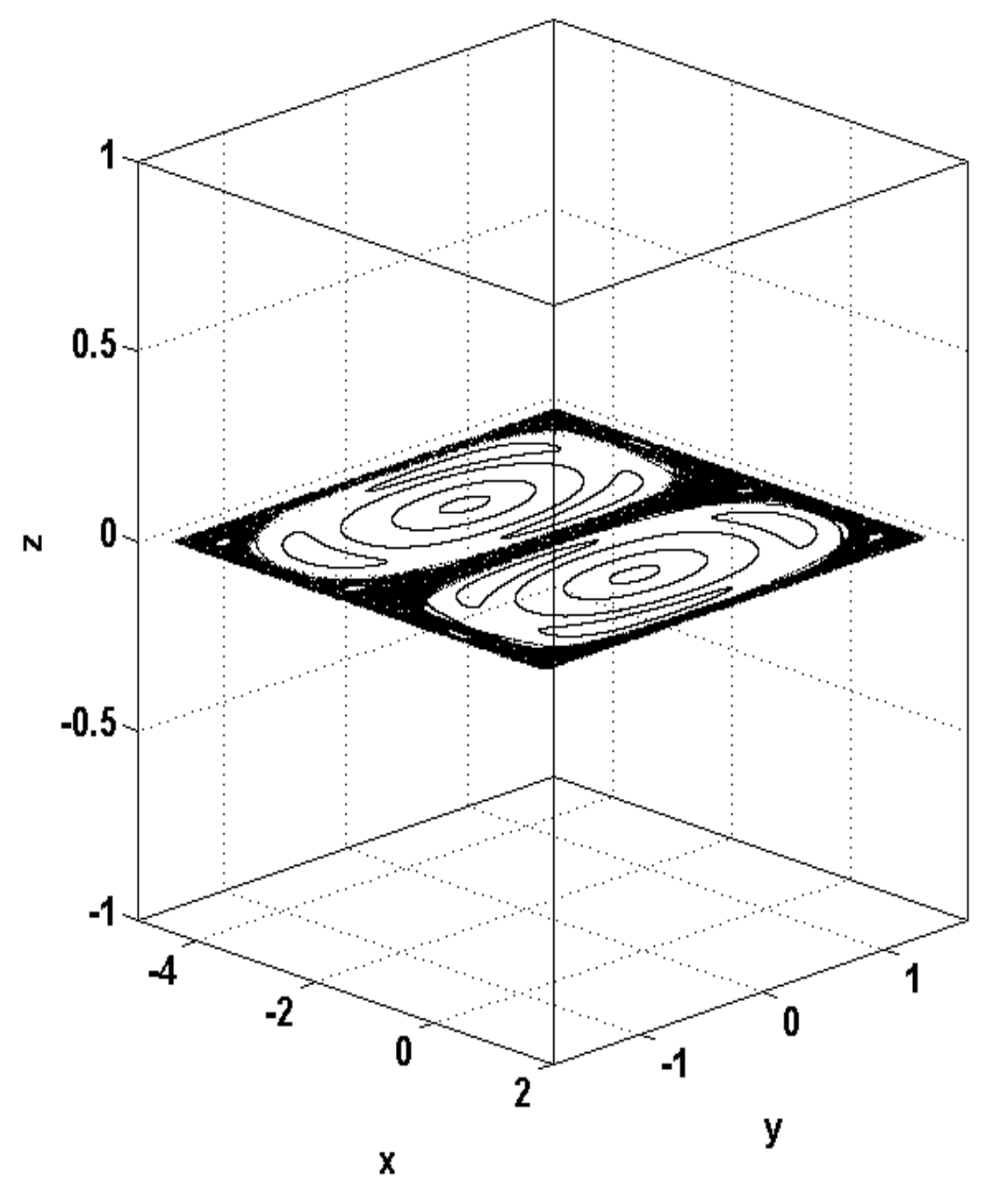}
\end{center}
\caption{Poincar\'e section  for the two-dimensional chaotic case with $\epsilon_0=0.10$, $\epsilon_1=0$ and $\omega=2.00$.}
\label{fig11b}
\end{figure}

The slightly unsteady two-dimensional case, corresponding to $0<\epsilon_0 \ll 1$ and $\epsilon_1 =0$ is characterized by the evolution of  a slow variable (perturbed action) and a fast variable (perturbed angle). The dynamical system expressed in the variables $(\psi,z,\chi)$  becomes
\begin{eqnarray*}
\dot{\psi}&=&\epsilon_0 \sin\omega t ~G_0\left(\psi,\chi\right),\\
\dot{z}&=&0,\\
\dot{\chi}&=&\Omega(\psi)+\epsilon_0 I_0\left(\psi,\chi\right),
\end{eqnarray*}
where $G_0\left(\psi,\chi\right)=\sin 2y$ and $I_0$ is $2\pi$ periodic in $\chi$.
The time perturbation enables the addition of a degree of freedom to the system, making the apparition of chaotic transport possible.
In such perturbed cases, the  phase space (real space in this case) dynamics is generally characterized by
a mixed structure: a regular region in the center of the half-cells and a chaotic region around
and between them, as shown in Fig.~\ref{fig11b}. Tracer particles within the vortices (localized around
the position of the elliptic fixed points present in the non-chaotic case) remain confined there without
being diffused to the other cells, while those in the chaotic sea (localized around the position
of the heteroclinic orbits present in the steady case) diffuse to the other cells through chaotic
transport. Such  a structure is relatively simple and well understood with the Kolmogorov \cite{Kolmogorov:54}, Arnold \cite{Arnold:63} and Moser \cite{Moser:62} (K.A.M.) theory. Indeed, the KAM theorem precisely identifies which invariant quasi-periodic tori (acting as impermeable barriers to transport) are simply deformed and
survive under the action of a weakly nonlinear perturbation from the tori which get destroyed
and subsequently become chaotic. 
\subsection{Three-dimensional chaotic case}
The addition of even a small perturbation in the third dimension can make
the structure of the flow much more complex. 
The number of slow variables and equivalently the number of time scales is crucial for the transport characteristics.
When $0< \epsilon_{0,1}\ll 1$ one has an action-action-angle perturbed system, i.e.~a system described by two slow variables  and one fast variable:
\begin{eqnarray*}
\dot{\psi}&=&\epsilon_0 \sin\omega t ~G_0\left(\psi,\chi\right)-2\epsilon_1\sin z ~\psi ~G_1\left(\psi,\chi\right),\\ 
\dot{z}&=&2\epsilon_1\cos z ~H\left(\psi,\chi\right),\\
\dot{\chi}&=&\Omega(\psi)+\epsilon_0 I_0\left(\psi,\chi\right)+\epsilon_1 I_1\left(\psi,z,\chi\right),
\end{eqnarray*}
where $H\left(\psi,\chi\right)=\cos 2y + \cos 2y$, $G_1\left(\psi,\chi\right)=\sin^2 x + \sin^2 y$ and $I_1$ is $2\pi$ periodic in $\chi$.

In such cases, behavior such as  diffusion \cite{CambridgeJournals:379074, springerlink:10.1007/BF01026490, PhysRevLett611799, PhysRevLett753669, Igor200151, VNM:2006, VWG:2007, PRE780263} are present and may generate complete chaotic mixing at very long times, 
unlike action-angle-angle perturbed flows in which surviving KAM-like regular tori act as barrier to transport. In oceanography, understanding how the effect of a component in the third dimension, which is small and present only during a relatively short time, can influence the transport is of crucial importance. 
Consequently, we focus on the difference between short- and long-time transport structures in the weakly three-dimensional perturbed case, i.e.~$0< \epsilon_{0}\ll \epsilon_{1} \ll 1$.
\section{\label{Long_structure} Long-time transport structures}
\subsection{Qualitative observations}
\begin{figure}[ht]
\begin{center}
\includegraphics[scale=.70]{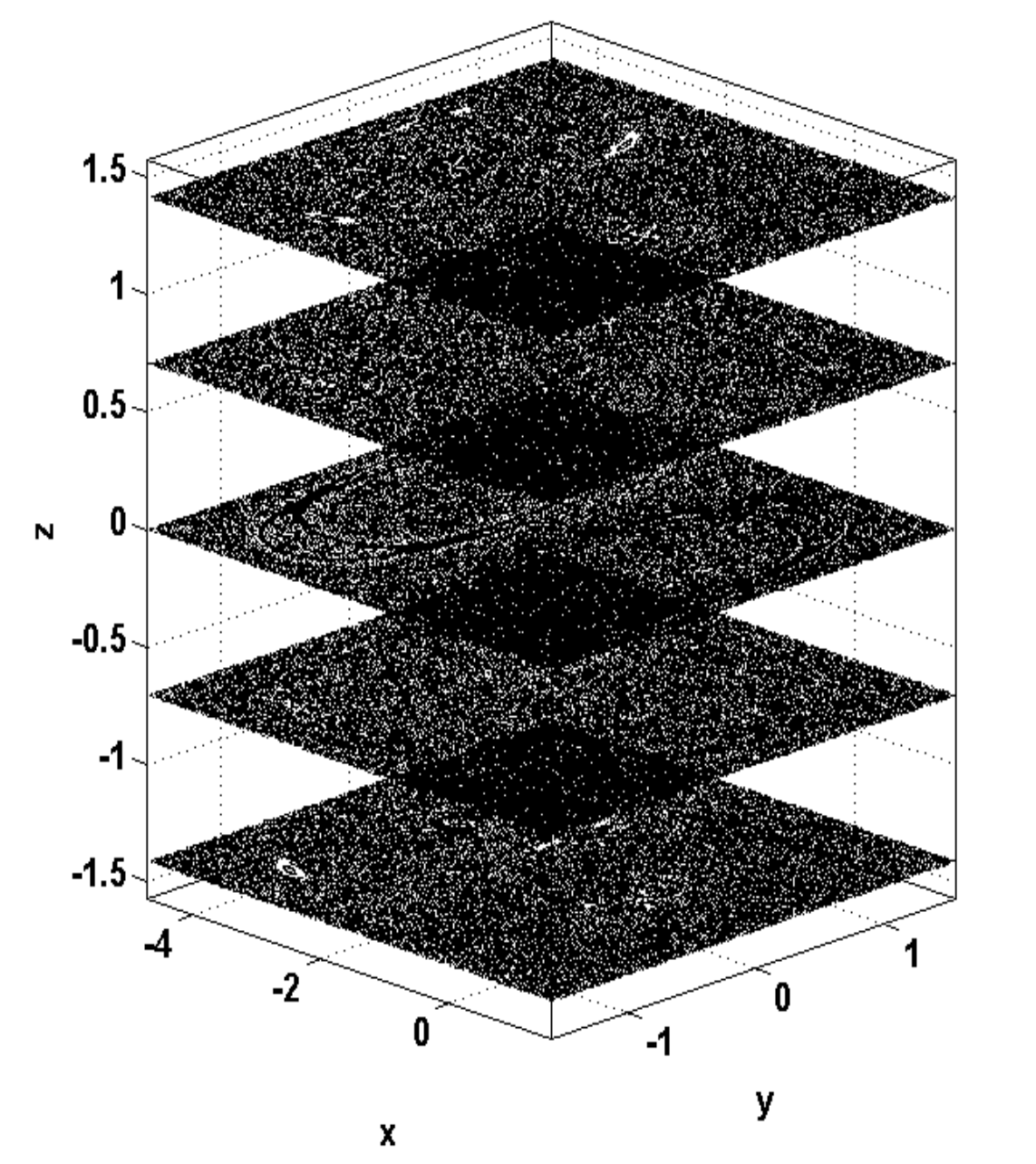}
\end{center}
\caption{Horizontal double Poincar\'e sections  (or Liouvillian sections) along the vertical axis for a single trajectory with parameter values $\epsilon_0=0.100$, $\epsilon_1=0.008$ and $\omega=2.000$.}
\label{fig3}
\end{figure}

The classical approach in dynamical systems theory consists of studying the long time qualitative behavior of all possible trajectories.
We do this  first by using one of the  most common tools, Poincar\'e sections.
Figures~\ref{fig3} displays the double Poincar\'e sections  (or Liouvillian sections, i.e.~two-dimensional projections, using a combination of a stroboscopic map and a plane section) of the unsteady three dimensional  flow (see Eqs.~\ref{VF1}-\ref{VF3}). Specifically, the Liouvillian sections considered here are
 the intersections of the trajectories at every period $2\pi/\omega$ with the planes $z=n\pi/6$, where $n=-2,-1,0,1,2$. \\
Each generic cell consists of  a chaotic mixing region covering practically the whole space. At this long time (more than $1000$ periods have been computed) the trajectories seem to wander chaotically everywhere throughout the space without any  apparent structure.
\subsection{\label{Lyapunov} Quantitative observations}
In order to gain further insight into the long-time transport structures within the channel, we compute the Lyapunov exponents map.
The technique consists of associating a Lyapunov exponent $\lambda$ with an initial condition ${\bm X}_{0}=(x_0,y_0,t=0)$.\\
First, we consider the time evolution of the Jacobian $ J^t\left(x,y\right)$ given by the tangent flow and the matrix of variations ${\bm \nabla}\bm{V}_{{\bm \lambda}}$ as
\begin{equation}
\frac{dJ^t}{dt}={\bm \nabla}{\bm V_{{\bm\lambda}}}\left(x,y,t\right) J^t,\\ 
\label{tangent flow}
\end{equation}
where  $J^0=I$ is the two-dimensional identity matrix. 
The Lyapunov map  is then defined as 
\begin{equation}
\lambda\left({\bm X}_{0}\right)=\lim_{\tau \to +\infty}\,\frac{1}{\tau}\ln\left|\gamma_{max}\left({\bm X}_{0}\right)\right|,
\label{Lyaponuv_profile}
\end{equation}
where $\gamma_{max}$ is the largest  eigenvalue (in norm) of the Jacobian $J^{\tau}$.

The Lyapunov map ${\bm X}_0\rightarrow \lambda\left({\bm X}_0\right)$  allows us to distinguish between the initial conditions leading to the presence of regular (non-mixing) islands, i.e. regions associated with zero Lyapunov exponent values, and the initial conditions  leading to chaotic mixing characterized by positive Lyapunov exponent value. 
The structures in phase space are then easily identified and the relative sizes of the regular (non-mixing) islands determined. This tool can be used to determine not only the phase space structures, but also quantify the degree of the mixing produced. A large Lyapunov exponent indicates strong stretching and folding, which is the archetypal mechanism of chaotic mixing. 
Due to the symmetry of the system, the domain of the Lyapunov exponent map reported below has been chosen as a periodic cell of the channel, i.e.~$\left\{-3\pi/2<x<\pi/2,-\pi/2<y<\pi/2,-\pi/2<z<\pi/2\right\}$.
\begin{figure}[ht]
\begin{center}
\includegraphics[scale=.70]{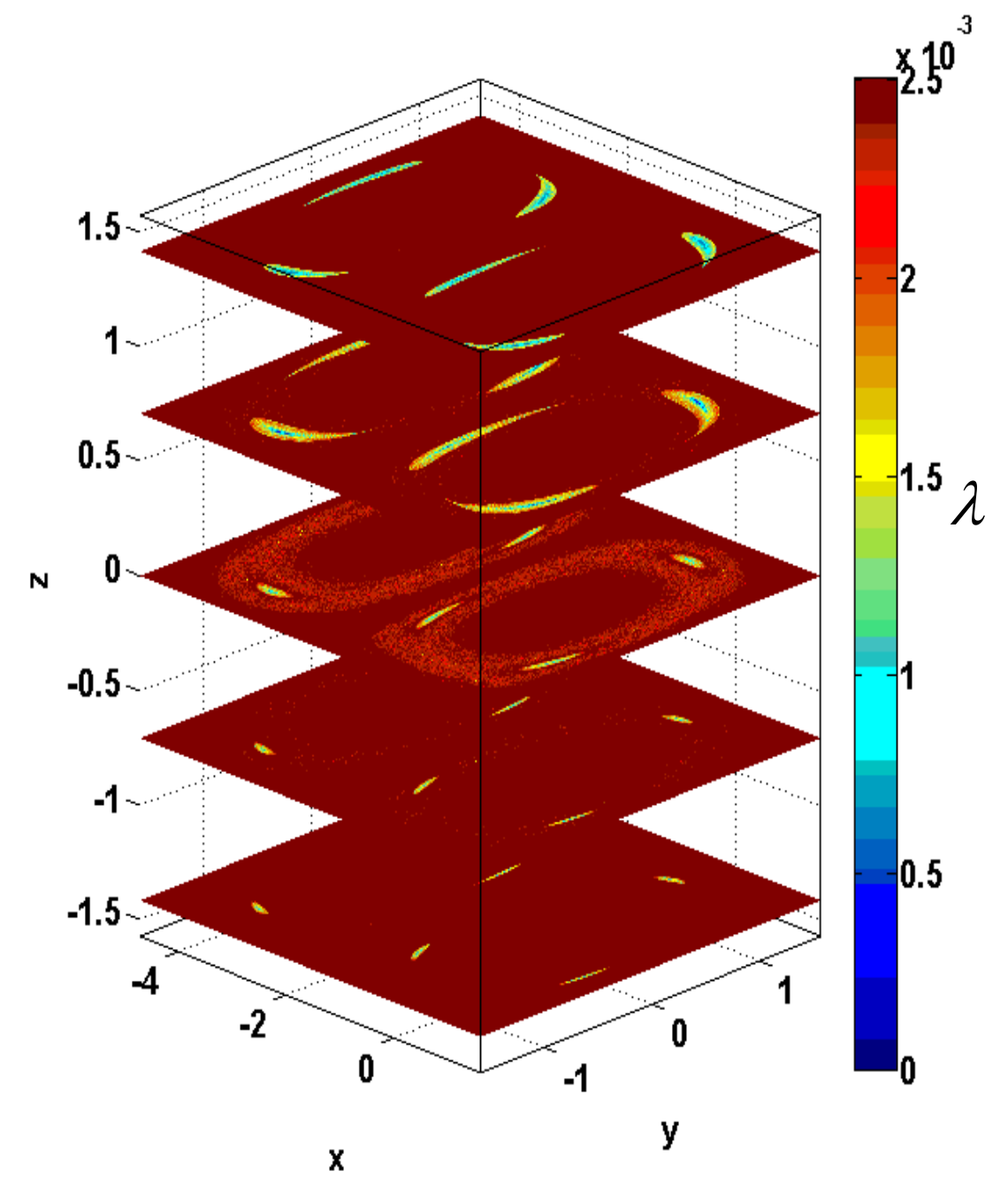}
\end{center}
\caption{Lyapunov exponent field on the planes $z=n\pi/6$, with $n=-2,-1,0,1,2$.}
\label{fig4}
\end{figure}
Figure \ref{fig4} shows horizontal sections of the Lyapunov field  at long time for parameter values $\epsilon_0 = 0.10$, $\epsilon_1=0.08$ and $\omega=2.00$. 
It is clear that in Fig.~\ref{fig4}, the chaotic transport is extremely well spread throughout the all cell space 
with a near-uniform distribution of the chaos intensity. 
\section{\label{Short_structure}Short-time transport structures}
\subsection{Short-time transport structure effects}
\subsubsection{Qualitative observations}
\begin{figure*}[ht]
\begin{center}
\includegraphics[scale=.8]{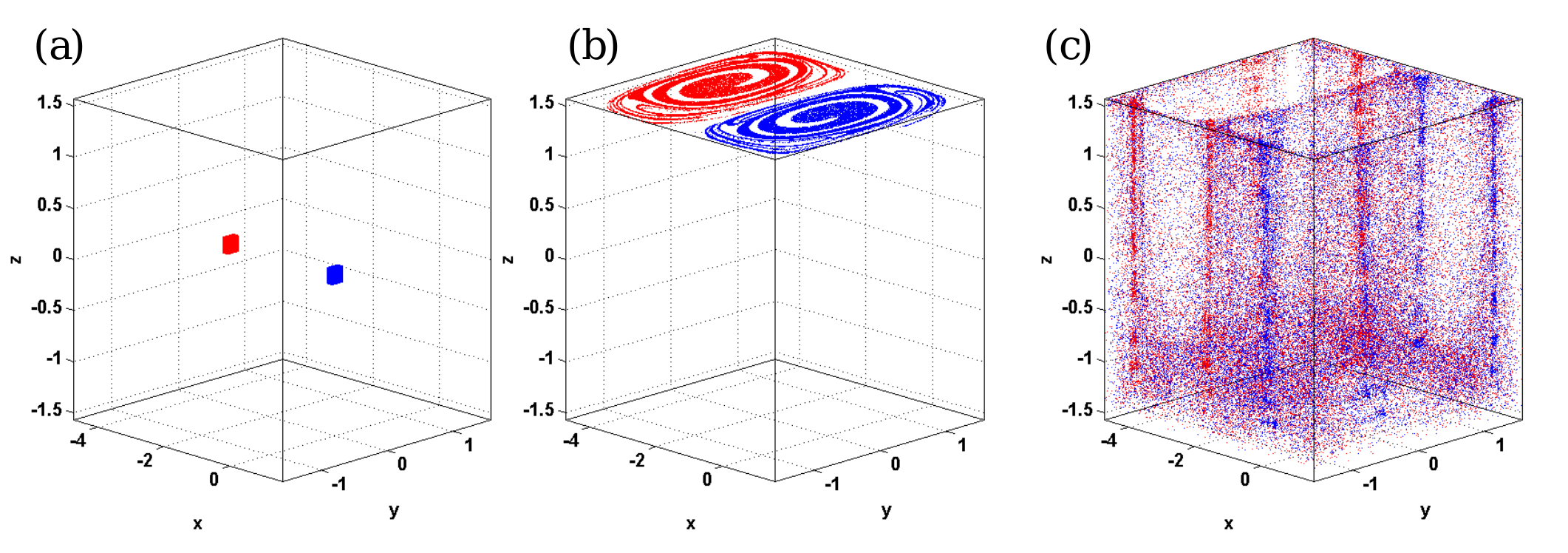}
\end{center}
\caption{Dynamics of two sets of $100,000$ advected particles at time $t=0,150\pi/\omega,400\pi/\omega$ (a-c), for oscillation amplitude $\epsilon_0=0.100$, vertical flow strength $\epsilon_1=0.008$ and  frequency $\omega=2.00$.}
\label{fig5}
\end{figure*}

First, we look at the transport behavior at short time by simply computing the time evolution of advected particles carried by the velocity field (see Eqs.~\ref{VF1}-\ref{VF3}).
Figure~\ref{fig5} shows how two sets of  advected particles, each initially concentrated at the center of the two half-cells, are transported during few hundreds of the forcing period $\omega$.
It is clear that depending on the location inside the cell, particles seem to be transported either \emph{regularly} 
or \emph{chaotically}, in contrast with the widespread chaotic transport present at long time. The particles stay confined around the vertical axis and  are regularly spread out  near the top wall without displaying any sign of sensitivity to initial condition. However, near and along the vertical walls of the half-cell, the particles are heavily deformed by a mechanism of successive stretching and folding.
\subsubsection{Quantitative observations}
The results from Sec.~\ref{Long_structure}  seem in contradiction with those presented here, depicting  Poincar\'e sections and Lyapunov field implying a well spread-out chaotic transport throughout the fluid domain (see Figs.~\ref{fig3}-\ref{fig4}). In order to gain more insight into this apparent contradiction, we have decided to quantify both the vertical and horizontal chaotic transport. 

A good way to quantify the degree of mixing or chaotic transport as a function of spatial location is by determining the mixing index $M$  through the box counting method (see Ref.~\cite{Stremler:08}). This technique offers the advantage of being relatively easy to implement, fast and rather cheap in computing power.
For this, we follow $N_p$ advected particles and divide the domain into $N_x \times N_y \times N_z$ boxes or cells.  At each time, the number of particles $n_i$ is computed in each box, and from this the fraction of the total number of particles, or particle rate $r_i$.  
Given the number of particles $n_i$ inside each box $i$, the computation is performed as follows.
\begin{eqnarray}
r_i&=&\frac{n_i}{n_p}~~\mbox{ if } n_i<n_p,\\
\nonumber
r_i&=&1~~~~~\mbox{ if } n_i\geq n_p,
\nonumber
\end{eqnarray}
where $n_p$ is the average number of advected particles, i.e.~$n_p=N_p/(N_x N_y N_z)$.
After computing the fraction of particles in each box and at each time $t$, the time evolution of the mixing index $M(t)$ is calculated by taking the average over all the boxes, i.e.
\begin{equation}
M(t)=\frac{1}{N_x N_y N_z}\sum_{i=1}^{N_x N_y N_z}r_i(t).
\nonumber
\end{equation}
A mixing index converging towards zero ($\lim_{t \to +\infty} M(t)=0$) indicates an extremely weak mixing process, while a mixing index converging towards one ($\lim_{t \to +\infty} M(t)=1$) corresponds to a perfect mixing process.
\begin{figure*}[ht]
\begin{center}
\includegraphics[scale=.70]{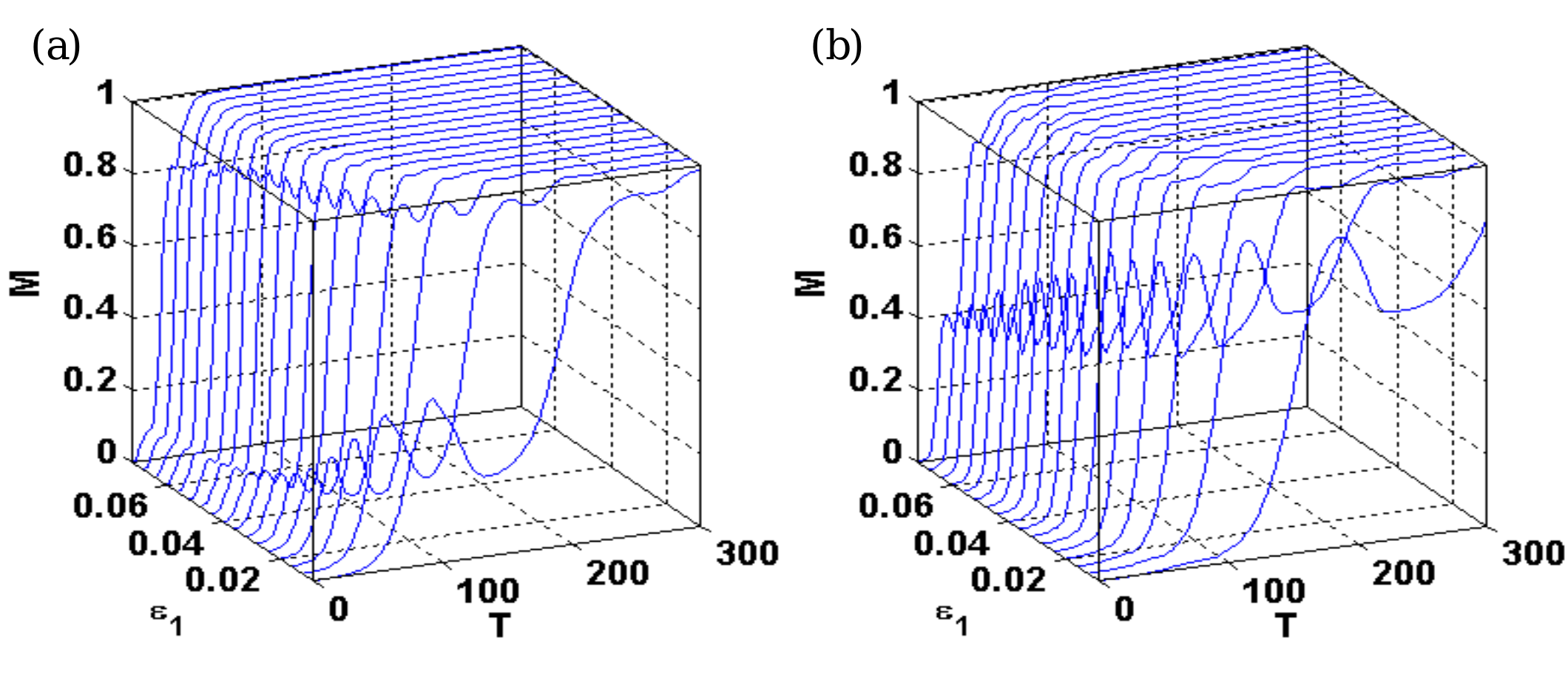}
\end{center}
\caption{a) Horizontal ($N_x \times N_y \times N_z= 3200 \times 3200 \times 1$) and b) vertical ($N_x \times N_y \times N_z= 3200 \times 1 \times 3200$)  mixing index as a function of the number of period $T=2\pi/\omega$ and vertical flow strength $\epsilon_1$ for $N_p=200,000$ advected particles.}
\label{fig6}
\end{figure*}
Figures~\ref{fig6} illustrates the mixing index $M$ in the horizontal (i.e. $N_x \times N_y \times N_z= 3200 \times 3200 \times 1$) and vertical (i.e. $N_x \times N_y \times N_z= 3200 \times 1 \times 3200$) direction  versus the time $t$ expressed in terms of the number of periods $T=2\pi/\omega$ and versus the vertical flow strength $\epsilon_1$. 

The time evolution of $M$ both for the horizontal and vertical transport shows a very interesting behavior. $M$ starts from a very small value then increases in an oscillatory fashion (instead of increasing monotonously) to reach, at longer time, a plateau at $M=1$ indicating complete mixing.
Such a behavior is in perfect accordance with the results seen up to here. 
The fact that the mixing index of both the vertical and horizontal mixing reaches at long time (i.e.~$t>600\pi/\omega$) the limit $M=1$ corroborates the qualitative (Poincar\'e sections in Fig.~\ref{fig3}) and quantitative (Lyapunov exponent field in Fig.~\ref{fig4}) results showing that the long-time chaotic transport is complete.
More importantly, the mixing index oscillations present at short times (representing alternations between horizontal and vertical chaotic transport) clearly reveal  specific structures at short times. 
The effects of these short-time structure is qualitatively observed through the time evolution of dense sets of advected particles that are transported regularly near the horizontal walls and chaotically near  the vertical walls.

Figure~\ref{fig6} shows also that as $\epsilon_1$ (i.e.~the vertical flow strength) is increased, the amplitude of  these oscillations diminish and vanish around $\epsilon_1\approx 0.1=\epsilon_0$. Such results, clearly indicate that this specific short-time structure is present only when the vertical flow is weak, i.e. $\epsilon_1<\epsilon_0$ and appears as some kind of remnant of the two-dimensional time dependent system $\epsilon_1=0$ (see Fig.~\ref{fig11b}).     

Now that we have seen that the presence of a short time structure characterized by the combination of regular transport (around the  vertical axis and near the horizontal walls) and chaotic transport (near the vertical walls of the half-cell), the next task  is  to  reveal and identify these structures.
\subsection{Identification of the short time transport structures}
\subsubsection{Short time Lyapunov exponent field}
As stated in Sec.~\ref{Lyapunov}, the long time structures in phase space can  easily be identified with the Lyapunov exponent field at infinite time.
Using the same approach, the short time structure can be revealed with the Finite Time Lyapunov Exponent (FTLE) field. 

\begin{figure*}[ht]
\begin{center}
\includegraphics[scale=.70]{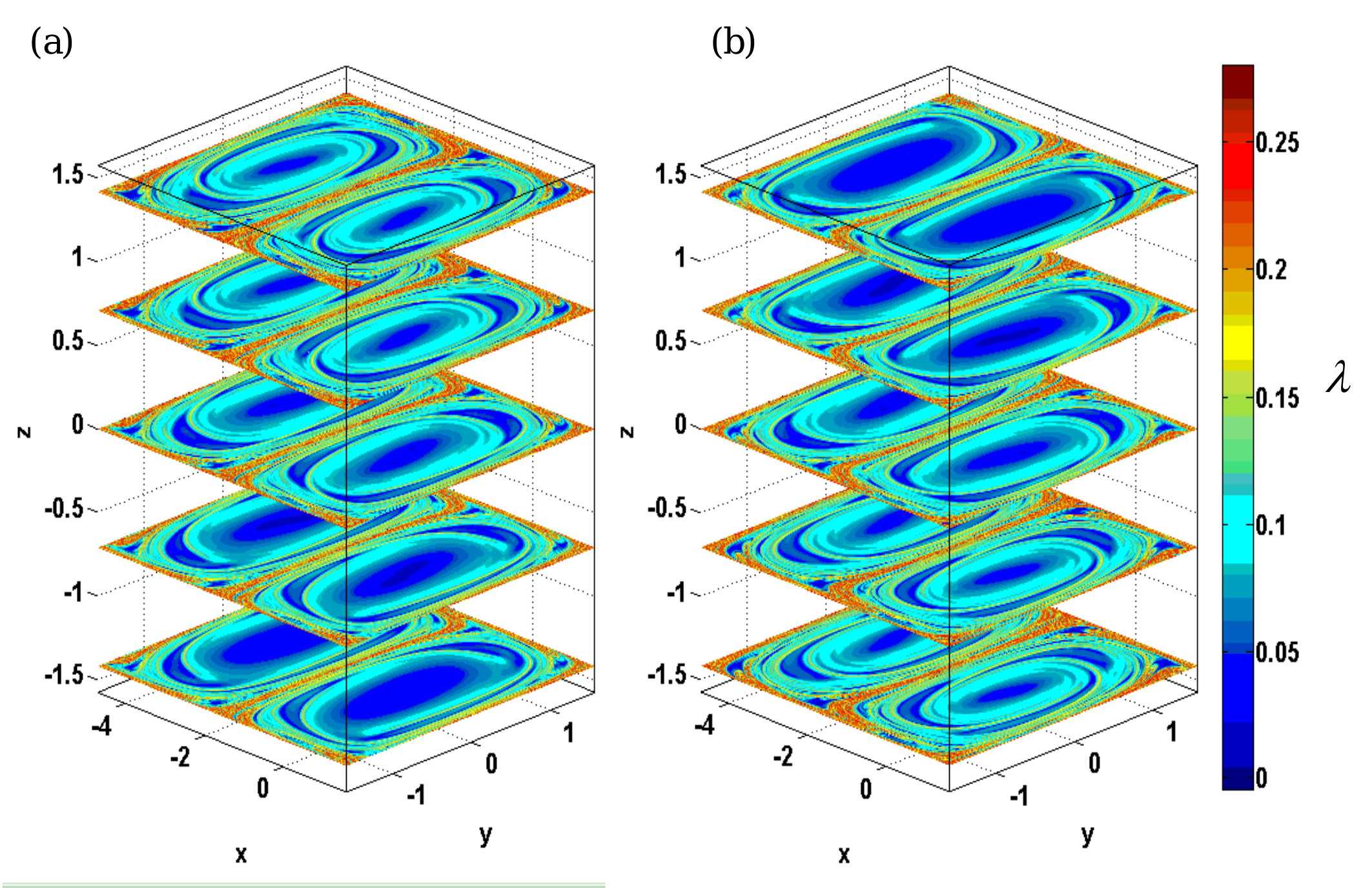}
\end{center}
\caption{a) Forward and b) backward finite time Lyapunov exponent map on horizontal sections at $\tau=\pm 20\pi/\omega$ with parameter values $\epsilon_0=0.100$, $\epsilon_1=.008$ and $\omega=2.00$.}
\label{fig7}
\end{figure*}

Figure~\ref{fig7} shows FTLE map ${\bm X}_0\rightarrow \lambda\left({\bm X}_0,\tau\right)$ on horizontal sections for time $\tau=\pm20\pi/\omega$.
FTLE maps for both positive and negative time have been  computing to provide information about the forward and backward dynamics.
In Fig.~\ref{fig7}, we distinguish two regions: a region of weak chaotic transport (cold color) located around the vertical axis joining the two horizontal walls and a region of strong chaotic transport (hot color) near the vertical walls. 
Once again these results corroborate the existence of a short time structure transporting regularly from bottom to  top walls via the vertical axis and chaotically near the vertical walls.
\subsubsection{Lagrangian coherent structure}
Recently, dynamically active barriers to transport labeled as Lagrangian Coherent Structures (LCS) have been revealed and studied in fluid flows \cite{haller:99,Haller2000352,G2001248,haller:3365}. These structures are now seen to be crucial in understanding  transport phenomena notably in time-dependent systems including oceanic\cite{Lekien:2005,Irina:2010}, atmospheric\cite{tang:017502} and physiological flows\cite{shadden:017512}.
These structures divide the fluid into dynamically distinct regions,  revealing geometry  hidden in the velocity field or trajectories of the system.
These attractive and repulsive Lagrangian coherent structures can be defined as the ridges of the finite time Lyapunov exponent  map calculated  backward and forward in time.

\begin{figure}[ht]
\begin{center}
\includegraphics[scale=.70]{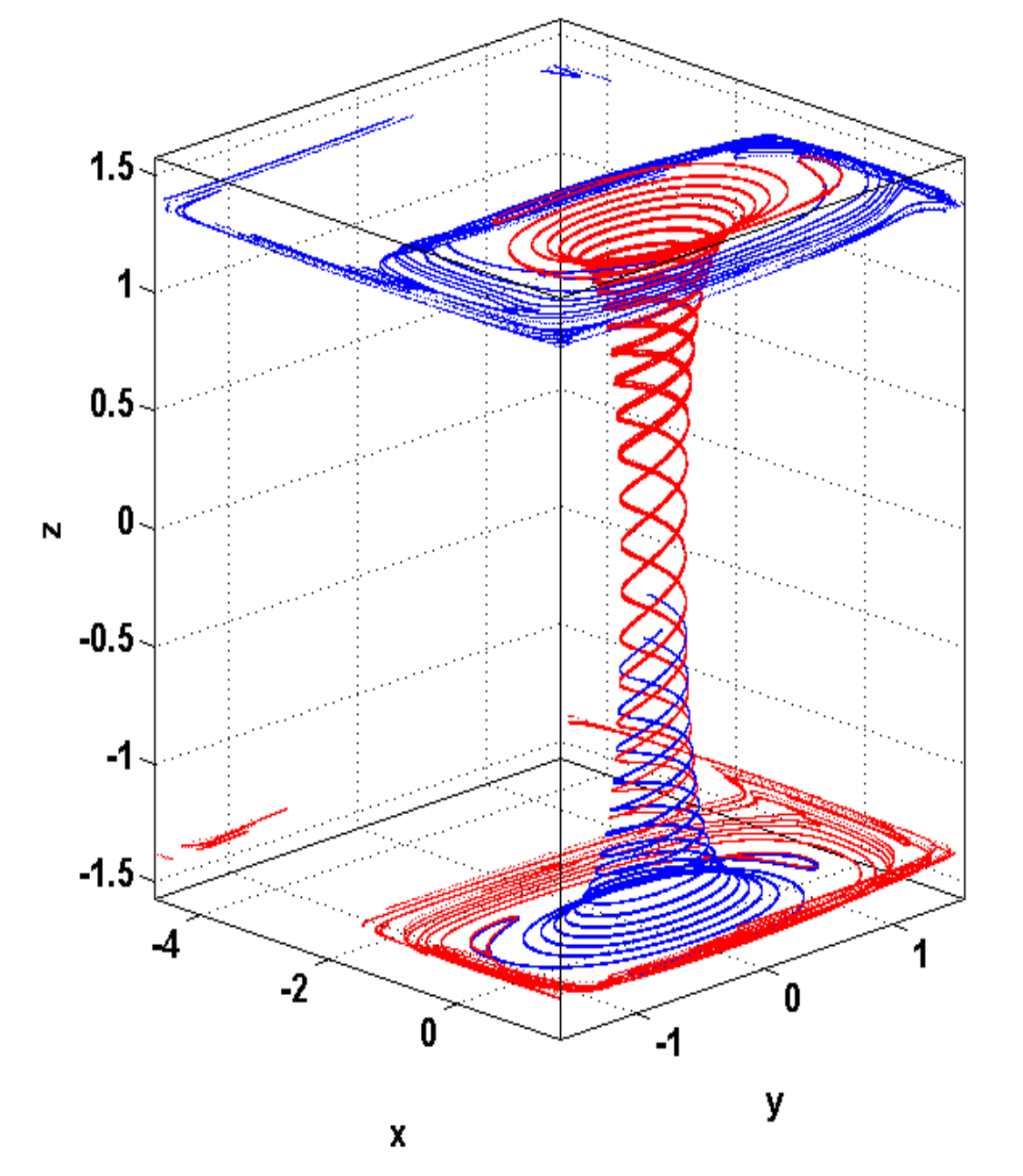}
\end{center}
\caption{Attractive (blue) and repulsive (red) Lagrangian coherent structures inside a generic half-cell. Only a part of the attractive and repulsive LCS extension have been represented for clarity.}
\label{fig8}
\end{figure}

Figure \ref{fig8} presents these LCS that have been computed from the ridges of the forward and backward FTLE map (see Figs. \ref{fig7}).
From these LCS that play the role of generalized attractive and repulsive manifolds (i.e. barriers to transport attracting or repulsing neighboring particles),
the short time transport structures can be unambiguously revealed.
From Fig.~\ref{fig8} we can see how the attractive (blue curve) and repulsive (red curve) LCS wind and tangle together. The attractive and repulsive LCS smoothly join together and wind around a surface starting from the bottom wall, passing around the vertical axis and finishing near the top wall. The fact that the attractive and repulsing LCS are smoothly joined  and formed one curve ensures the presence of regular transport on this region and at this short time. Near the top wall  and close to the vertical wall three hyperbolic structures separate the attractive and repulsive LCS (see Fig.~\ref{fig8}).  
The attractive (or repulsive) LCS coming from the top (or bottom) wall begin to wander and tangle along the neighboring of the vertical walls, intersecting transversely the repulsive (attractive), thus indicating the presence of chaotic transport.  
\section{Conclusion}
In this paper, we have exposed how the short-time  differs from the long-time Lagrangian transport behavior for a  three-dimensional time-dependent fluid model. More precisely, the model features are characteristic of oceanic flows and can be completely described by a set of three variables, one extremely slow, one slow and one fast.
Contrary to  most approaches that focus only on transport behavior at infinite time or  averaged over time, we have turned our attention to the transport structures at short time.
We have shown both qualitatively (using snapshots of the time evolution of advected particles) and quantitatively (using horizontal and vertical mixing index) 
the presence of  specific short-time transport structures. In addition, using Lagrangian Coherent Structures we were able to identify and characterize these short-time transport structures.
These short-time structures show strong and fast chaotic transport along vertical half-cell boundaries as well as seemingly regular transport around the half-cell vertical axis and near the top and bottom half-cell boundaries. These short-time structures contrast with the long-time behavior showing a spread-out chaotic transport throughout the cell.
This knowledge of  short-time transport structures is of great
interest in numerous concrete applications, particularly for oceanic flows where simulations or observations are available only over a short time.
\section*{Acknowledgments}
This research was funded by the ONR MURI Dynamical Systems Theory and Lagrangian Data Assimilation in 3D+1 Geophysical Fluid Dynamics.
\bibliographystyle{apsrmp4-1}
\bibliography{biblioVIII}
\end{document}